\documentclass{article}
\usepackage[preprint]{spconfa4}
\usepackage{amsmath,graphicx,booktabs}
\usepackage{color}
\usepackage{cite}
\usepackage{amssymb,amsfonts,url,amsmath}
\usepackage{algorithmic}
\usepackage{textcomp}
\usepackage[rgb]{xcolor}
\usepackage{float}
\usepackage{placeins}
\newcommand{\stz}{\rule{0mm}{2.3ex}}

\usepackage{bm}
\usepackage{color,soul,cite}
\usepackage{MnSymbol}
\usepackage{wasysym}
\usepackage{multirow}
\usepackage{makecell}
\usepackage{colortbl,hhline}
\usepackage[super]{nth}
\usepackage{subcaption}
\title{Does a PESQNet (Loss) Require a Clean Reference Input?\\The Original PESQ Does, But ACR Listening Tests Don't
}
\name{Ziyi Xu, Maximilian Strake, Tim Fingscheidt}
\address{Institute for Communications Technology, Technische Universit{\"a}t Braunschweig, Germany\\$\left \{ \text{ziyi.xu, m.strake, t.fingscheidt} \right \}$@tu-bs.de}
\begin{document}
\ninept
\maketitle
\begin{abstract}
Perceptual evaluation of speech quality (PESQ) requires a clean speech reference as input, but predicts the results from (reference-free) absolute category rating (ACR) tests. In this work, we train a fully convolutional recurrent neural network ({\tt FCRN}) as deep noise suppression (DNS) model, with either a {\it non-intrusive} or an {\it intrusive} {\tt PESQNet}, where only the latter has access to a clean speech reference. The {\tt PESQNet} is used as a mediator providing a perceptual loss during the DNS training to maximize the PESQ score of the enhanced speech signal. For the {\it intrusive} {\tt PESQNet}, we investigate two topologies, called early-fusion (EF) and middle-fusion (MF) {\tt PESQNet}, and compare to the {\it non-intrusive} {\tt PESQNet} to evaluate and to quantify the benefits of employing a clean speech reference input during DNS training. Detailed analyses show that the DNS trained with the MF-intrusive {\tt PESQNet} outperforms the Interspeech 2021 DNS Challenge baseline and the same DNS trained with an MSE loss by $0.23$ and $0.12$ PESQ points, respectively. Furthermore, we can show that only marginal benefits are obtained compared to the DNS trained with the {\it non-intrusive} {\tt PESQNet}. Therefore, as ACR listening tests, the {\tt PESQNet} does not necessarily require a clean speech reference input, opening the possibility of using real data for DNS training.
\end{abstract}
\vspace*{-3mm}
\section{Introduction}
\vspace*{-2mm}
Speech quality is an important factor in evaluating speech enhancement algorithms, typically measured through subjective listening tests or by instrumental measurements such as perceptual evaluation of speech quality (PESQ) \cite{ITUT_pesq_wb_corri} or perceptual objective listening quality assessment (POLQA)\cite{ITUT_polqa_2018}. However, obtaining listening scores through subjective listening tests can be time-consuming and expensive. Although the software requires a clean speech reference signal, PESQ \cite{ITUT_pesq_wb_corri} is designed to predict absolute category rating (ACR) listener scores and is a widely used instrumental measure.

Speech enhancement algorithms employing deep neural networks (DNNs) have attracted a lot of research attention in recent years \cite{williamson2016complex,zhao2018convolutionalrecurrent,wang2018supervised,elshamy2018dnn,strake2019separated,strake2020fully,braun2020data,strake2020DNS}, and are subsumed under the term deep noise suppression (DNS). During the training process, most of the DNS architectures are trained with a mean squared error (MSE) loss, which does not guarantee good human perceptual quality of the enhanced speech signal \cite{liu2017perceptually,kolbcek2018monaural,zhang2018training,martin2018deep,fu2018end,zhao2019perceptual,fu2019learning,braun2021consolidated,xu2021inter,xu2021deepT}. To mitigate this problem, a straightforward solution could be to directly adapt PESQ as a loss function. However, the original PESQ implements a non-differentiable function, which cannot directly be used as an optimization criterion for gradient-based deep learning. {M}art{\'\i}n-{D}o{\~n}as et al.\ proposed an approximated differentiable PESQ formulation as the optimization criterion \cite{martin2018deep}, however, not exploiting the full potential of PESQ as a loss, as other simple psychoacoustic losses turned out to perform superior on the PESQ metric \cite{zhao2019perceptual}. Fu et al.\ \cite{fu2019learning} trained an end-to-end neural network (so-called {\tt Quality-Net}) to approximate the PESQ function. Afterwards, the trained {\tt Quality-Net} is fixed to estimate the PESQ scores of the enhanced speech, serving as a differentiable PESQ loss for the training of a DNS model. However, as reported by the authors of \cite{fu2019learning}, the gradient obtained from the fixed {\tt Quality-Net} after training for several minibatches, leading to the {\tt Quality-Net} being fooled by the updated DNS model: Estimated PESQ scores increase while true PESQ scores decrease. Please note that like the original PESQ \cite{ITUT_pesq_wb_corri}, both \cite{martin2018deep,fu2019learning} require a clean speech reference signal.

In our recent works \cite{xu2021inter,xu2021deepT}, we proposed an end-to-end {\it non-intrusive} {\tt PESQNet}, which is adapted from a speech emotion recognition DNN proposed in \cite{Meyer2021}, to model the PESQ function without knowing the corresponding clean speech (like human raters in ACR listening tests). Subsequently, the trained {\tt PESQNet} is employed as a mediator during the training of a DNS model aiming at maximizing the PESQ score of the enhanced speech signal. In \cite{xu2021deepT}, we proposed a successful training protocol to train the DNS and the {\tt PESQNet} alternatively on an {\it epoch} level to keep the {\tt PESQNet} up-to-date. Therefore, the {\tt PESQNet} can always adapt to the current updated DNS model, which solves the problems addressed in \cite{fu2019learning}. Only for a simple four-layer PESQ-estimating CNN (with two final fully-connected layers), Gamper et al.\ have shown that an extra clean speech reference input channel (``early fusion") is helpful \cite{gamper2019intrusive}. As PESQ performs middle fusion, it is an open question, whether a DNS model trained with a more powerful {\it intrusive} {\tt PESQNet} (early or middle fusion?) would perform better than trained with a powerful {\it non-intrusive} {\tt PESQNet}.

In this work, our contributions are threefold: First, we train a fully convolutional recurrent neural network ({\tt FCRN}) \cite{strake2020fully} as our DNS model with either an {\it intrusive} or a {\it non-intrusive} {\tt PESQNet} on the same training dataset constructed from the Interspeech 2021 DNS Challenge \cite{reddy2021interspeechIN}. Second, for the {\it intrusive} {\tt PESQNet} employing the clean speech reference input, we investigate two topologies, called early-fusion and middle-fusion {\tt PESQNet}, respectively. Finally, we perform an extensive analysis and experimental evaluation on the trained DNS models to evaluate the effects of using the additional clean speech reference and show that without any significant disadvantage a {\it non-intrusive} {\tt PESQNet} can be employed to optimize a DNS model for PESQ. This keeps the door open for training of DNS models on real data --- which, beyond {\tt PESQNet} \cite{xu2021inter,xu2021deepT}, was so far only shown with generative adversarial networks (GANs) \cite{pascual2017seganinter,pandey2018adversarial,li2020imetricgan}.

The rest of the paper is structured as follows: Section 2 introduces notations and the speech enhancement system. The PESQNet losses provided by {\it intrusive} or {\it non-intrusive} {\tt PESQNet} are presented in Section 3. We explain the experimental setup and discuss the results in Section 4, concluding the work in Section 5.
%\vspace*{-2mm}
\section{Signal Model and Notations}
\vspace*{-2mm}
\begin{figure}[t!]
	\centering
	\includegraphics[width=0.49\textwidth]{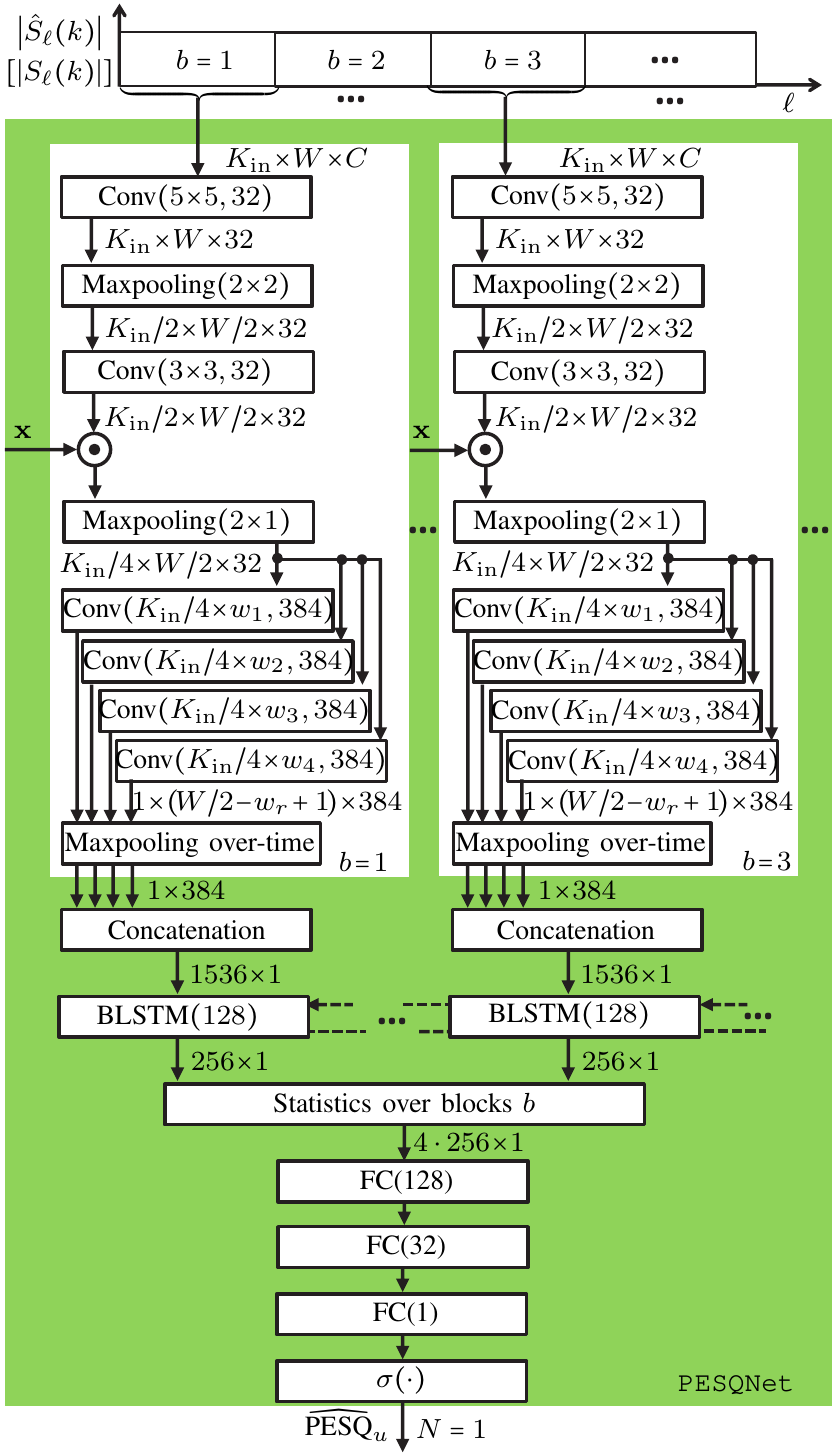}
	\caption{{\bf Employed} \texttt{\textbf{PESQNet}} as used in Fig.\,\ref{system3}. For the non-intrusive and the MF-intrusive {\tt PESQNet}, the number of input channels is set to $C\!=\!1$, while $C\!=\!2$ holds for the EF-intrusive {\tt PESQNet}. The element-wise factor $\mathbf{x}$ is set to an all-ones tensor for non-intrusive and EF-intrusive {\tt PESQNet}. For the MF-intrusive {\tt PESQNet}, $\mathbf{x}$ is computed from $\left|S_\ell(k)\right|\!-\!\left|\hat{S}_\ell(k)\right|$ as additional reference input.}
	\label{fig:PESQNet}
	\vspace*{-4mm}
\end{figure}
We assume the microphone mixture $y(n)$ to be constructed from the clean speech signal $s(n)$ reverberated by the room impulse response (RIR) $h(n)$, and disturbed by additive noise $d(n)$ as
\vspace*{-1mm}
\begin{equation} \label{micro_mixture}
\vspace*{-1mm}
y(n)=s(n)*h(n)+d(n)= s^\text{rev}(n)+d(n),
\end{equation}
with $s^\text{rev}(n)$ and $n$ being the reverberated clean speech signal and the discrete-time sample index, respectively, and $*$ denoting a convolution operation. Afterwards, all the signals are transformed to the discrete Fourier transform (DFT) domain:
\begin{equation} \label{micro_fft}
\vspace*{-1mm}
Y_\ell(k)=S^\text{rev}_\ell(k)+D_\ell(k),
\vspace*{-1mm}
\end{equation}
with frame index $\ell$, frequency bin index $k\!\in\!\mathcal{K}\!=\!\left \{0,1,\ldots,K\!-\!1\right \}$, and $K$ being the DFT size. Our employed DNS is the {\tt FCRN} from \cite{strake2020fully}, which delivers a magnitude-bounded complex mask $M_\ell\left(k \right )\in\mathbb{C}$, with $\left|M_\ell\left(k \right )\right|\in\left [ 0,1 \right ]$ for spectrum enhancement \cite{strake2020DNS}. Therefore, the enhanced speech spectrum is obtained by:
\begin{equation} \label{clean_speech_est}
\vspace*{-1mm}
\hat{S}_\ell\left (k \right )=Y_\ell(k)\cdot M_\ell\left(k \right ).
\vspace*{-1mm}
\end{equation}
Finally, the enhanced speech spectrum $\hat{S}_\ell\left (k \right )$ is subject to an inverse DFT (IDFT), followed by overlap add (OLA) to reconstruct the estimated signal $\hat{s}(n)$.
\vspace*{-1mm}
\section{PESQNet Loss --- Intrusive and Non-Intrusive}
\vspace*{-2mm}
In this work, we employ an end-to-end {\it intrusive} or {\it non-intrusive} {\tt PESQNet} to model ITU-T P862.2 PESQ \cite{ITUT_pesq_wb_corri}. The employed {\tt PESQNet} aims at estimating the PESQ score of an entire enhanced speech utterance in the DFT domain. Therefore, the {\tt PESQNet}'s estimation $\widehat{\text{PESQ}}_u$ for the utterance indexed with $u$ should be close to its ground truth $\text{PESQ}_u$ measured by ITU-T P.862.2 \cite{ITUT_pesq_wb_corri}. Thus, the ``PESQ loss" used for training the {\tt PESQNet} is (see Fig.\,\ref{system3}):
\begin{equation} \label{PESQNet}
J^\text{PESQ}_u\!=\left(\widehat{\text{PESQ}}_u-\text{PESQ}_u\right)^2.
\end{equation}
\subsection{Non-Intrusive \texttt{PESQNet}}
In our recent works \cite{xu2021inter,xu2021deepT}, we have proposed the {\it non-intrusive} {\tt PESQNet} as depicted in Fig.\,\ref{fig:PESQNet}. The input of the {\it non-intrusive} {\tt PESQNet} is the enhanced amplitude spectrogram $\left|\hat{S}_\ell(k)\right|$, with $\ell\!\in\!\mathcal{L}_u\!=\!\left \{1,2,\ldots,L_u\right \}$, and $L_u$ being the number of frames for an entire utterance $u$. Since the amplitude spectrogram $\left|\hat{S}_\ell(k)\right|$ is used as input, the number of input channels is set to $C=1$ in Fig.\,\ref{fig:PESQNet}. The input is then grouped into several blocks indexed with $b\in\mathcal{B}_u=\left\{1, 2 \ldots, B_u\right\}$. Each block has the same dimension $K_{\rm in}\!\times\! W\! \times\! 1$, with $K_{\rm in}$ and $W$ being the number of input frequency bins and frames per block, respectively. The convolutional layers are represented by Conv$(h\times w, f)$, with $f$ representing the number of filter kernels, and $(h\times w)$ being the kernel size. We employ maxpooling layers with two different kernels of size $(2\times1)$ and $(2\times2)$, respectively. The maxpooling-over-time layer and the subsequent concatenation aim to deliver a feature map with a fixed dimension to the bidirectional LSTM (BLSTM) layer with $128$ nodes. Afterwards, four statistics (average, standard deviation, minimum, and maximum) over blocks $b$ are applied to the BLSTM outputs. The fully-connected (FC) layer is denoted as FC$(N)$, with $N$ being the number of neurons. The singe-node output layer employs a gate function $\sigma(x)=3.6\cdot\text{sigmoid}(x)+1.04$ to limit the range of the estimated PESQ score between $1.04$ and $4.64$, as defined in the original PESQ \cite{ITUT_pesq_wb_corri}. Please note that in Fig.\,\ref{fig:PESQNet}, the element-wise factor $\mathbf{x}$ is an all-ones tensor for the {\it non-intrusive} {\tt PESQNet}.
\vspace*{-2mm}
\subsection{Intrusive \texttt{PESQNet}}
\vspace*{-1mm}
For the {\it intrusive} {\tt PESQNet}, we investigate two different topologies for employing the clean speech reference input, called early-fusion (EF) and middle-fusion (MF) {\tt PESQNet}, respectively. The idea of the EF-intrusive {\tt PESQNet} is very straightforward: the amplitude spectrograms of the enhanced speech $\left|\hat{S}_\ell(k)\right|$ and its corresponding clean speech reference $\left|S_\ell(k)\right|$ are employed as two separate channels for the input of the {\tt PESQNet}. Therefore, the number of input channels is set to $C=2$ in Fig.\,\ref{fig:PESQNet}. As with the {\it non-intrusive} {\tt PESQNet}, the input is grouped into several blocks with the same dimensions for parallel processing, as illustrated in Fig.\,\ref{fig:PESQNet}. For the EF-intrusive {\tt PESQNet}, the element-wise factor $\mathbf{x}$ is an all-ones tensor.

Inspired by the original PESQ \cite{ITUT_pesq_wb_corri}, we propose an MF-intrusive {\tt PESQNet}, which explicitly considers the {\it differences} between the degraded signal and its corresponding clean reference signal during the PESQ score estimation. Compared to the {\it non-intrusive} {\tt PESQNet}, we introduce an {\it additional separate} input branch, which explicitly uses the differences $\left|S_\ell(k)\right|\!-\!\left|\hat{S}_\ell(k)\right|$ as input. This separate differential-input branch has the same topology as the main branch (Fig.\,\ref{fig:PESQNet}) until reaching the element-wise multiplication, but employs a sigmoid activation to ensure $0\!\leq\!x\!\leq\!1$ for each element of tensor $\mathbf{x}=(x)$. Therefore, the additional input branch is used to control how much information from the original input of the enhanced speech spectrogram contributes to the final PESQ score estimation. 
%%%%%%%%%%%%%%%%%%%%%%%%%%%%%%%%%%5
\begin{figure}[t!]
	\centering
	\includegraphics[width=0.49\textwidth]{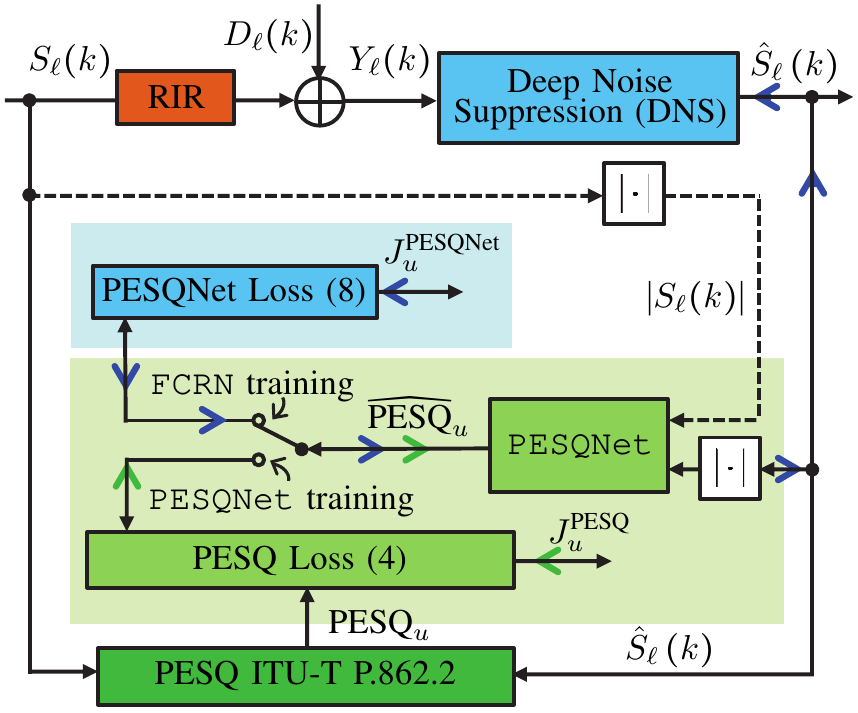}
	\caption{{\bf \texttt{PESQNet} and DNS joint training setup.} The DNS and the {\tt PESQNet} are trained alternately, controlled by the switch in upper and lower position, respectively. Colored arrows: gradient flow.}
	\label{system3}
	\vspace*{-3mm}
\end{figure}
\vspace*{-1mm}
\subsection{PESQNet Loss for DNS training}
\vspace*{-1mm}
Building upon our recent works \cite{xu2021inter,xu2021deepT}, we employ the proposed {\tt PESQNet} to control the fine-tuning on a pre-trained DNS model to further increase the perceptual quality of the enhanced speech signal. Therefore, our employed DNS model is pre-trained employing the utterance-wise loss function proposed in \cite{strake2020DNS} including two MSE loss terms. The first loss term aims at joint dereverberation and denoising by employing the clean speech spectrum $S_\ell(k)$ as target:
%%%%%%%%%%%%%%%%%%%%%%%%%%%%%%%%%%%%%%%%%%%
\begin{equation} \label{joint}
\vspace*{-1mm}
J^\text{joint}_u\!=\!\frac{1}{L_u\!\cdot\!K}\!\sum_{\ell\in\mathcal{L}_u}\sum_{k\in\mathcal{K}}\!\bigl|\hat{S}_\ell(k)\!-\!S_\ell(k)\bigr|^2,
\vspace*{-1mm}
\end{equation}
with $\mathcal{L}_u$ being the set of frame indices for an utterance indexed with $u$, and $L_u$ being its total number of frames. The second loss term only focuses on denoising by employing the reverberated clean speech spectrum $S^\text{rev}_\ell(k)$ as target:
\begin{equation} \label{noise}
%\vspace*{-1mm}
J^\text{noise}_u\!=\!\frac{1}{L_u\!\cdot\!K}\!\sum_{\ell\in\mathcal{L}_u}\sum_{k\in\mathcal{K}}\!\bigl|\hat{S}_\ell(k)\!-\!S^\text{rev}_\ell(k)\bigr|^2\!.
\vspace*{-1mm}
\end{equation}
Afterwards, the two loss terms \eqref{joint} and \eqref{noise} are combined into a joint loss function as:
\begin{equation} \label{MT}
J^\text{MSE}_u\!=\alpha\cdot J^\text{joint}_u+(1-\alpha)\cdot J^\text{noise}_u,
\end{equation}
with $\alpha=0.9$ being the weighting factor to control the dereverberation effect. Afterwards, the proposed {\tt PESQNet} is pre-trained employing the PESQ loss \eqref{PESQNet} with the enhanced speech signal obtained from the fixed pre-trained DNS.

In the fine-tuning stage shown in Fig.\,\ref{system3}, the pre-trained {\tt PESQNet} is applied to the output of the pre-trained DNS to estimate the PESQ scores of the enhanced speech. Thus, we can define a ``PESQNet loss" provided by the proposed {\tt PESQNet} to maximize the PESQ scores of the output enhanced speech signal as:
\begin{equation} \label{real}
J^\text{PESQNet}_u\!=\left(\widehat{\text{PESQ}}_u-\text{PESQ}_\text{max}\right)^2
\end{equation}
for utterance $u$, with $\text{PESQ}_\text{max}\!=\!4.64$, which is minimized during DNS fine-tunning. We adopt the successful joint training protocol proposed in \cite{xu2021deepT} to fine-tune the DNS and the {\tt PESQNet} alternatingly on an {\it epoch} level to keep the {\tt PESQNet} up-to-date, which is controlled by the switch in the upper and lower positions, as shown in Fig.\,\ref{system3}. The blue and green arrows indicate the gradient flow back-propagated for the DNS and {\tt PESQNet} training, respectively. Please note that in Fig.\,\ref{system3}, the dashed clean reference signal exists only for employing the {\it intrusive} {\tt PESQNet}.
%%%%%%%%%%%%%%%%%%%%%%%%%%%%%%%%%%%%%%%%%%%%%%%%%%%%%%%%%%%%%%%%%%%%%%%%%%%
\begin{table*}[!t]
	\centering
	\caption{\textbf{Instrumental quality results} on the \textbf{development set} $\mathcal{D}^\mathrm{dev}_\mathrm{DNS1}$. Best results are in {\bf bold} font, and the second best are \underline{underlined}.}
	\setlength\tabcolsep{4pt}
	\vspace*{-2mm}
	%\resizebox{0.98\linewidth}{!}{
	\begin{tabular}{ccc c c c c c c c}
		\hline
		&\multirow{2}{*}{Methods} & \multicolumn{4}{c}{\stz{Without reverb}}& \multicolumn{4}{c}{\stz{With reverb}}\\ \cmidrule(r){3-6} \cmidrule(r){7-10} %\hhline{~~--------}
		&    &  PESQ & DNSMOS  &  \stz{STOI}  &   $\Delta\text{SNR}_\text{seg}$[dB]   & PESQ & DNSMOS   & \stz{STOI} &  SRMR \\ \hhline{----------} 
		\multirow{4}{*}{}& \multicolumn{1}{l}{Noisy}             &  \stz{2.21} &  3.15  &  0.91   &  -  &  1.57  &  2.73  &0.56    &-  \\ \hhline{~~~~~~~~~~}
		& \multicolumn{1}{l}{\stz{DNS3 Baseline \cite{braun2020data}}}              &     3.15   &  3.64    &    0.94   &    {6.30} & \underline{1.68}   &  {\bf 3.18}   & \underline{0.62}   & 6.33\\ \hhline{~~~~~~~~~~}
		& \multicolumn{1}{l}{\stz{{\tt FCRN} \cite{strake2020DNS}}}  &      {3.37}   &  {3.82}   &    {\bf 0.96}   &    8.35 & {\bf 1.95} &  3.08    &  {\bf 0.63}  & {7.25}\\ \hhline{~~~~~~~~~~}
		&\multicolumn{1}{l}{\stz{{\tt FCRN}/{\tt PESQNet}, non-intrusive \cite{xu2021deepT}}}   &    \underline{3.45}  &  \underline{3.87}  & {\bf 0.96}  &    {8.48}  & {\bf 1.95}  &  \underline{3.13}   &  \underline{0.62}  & \underline{7.38} \\\hhline{----------}
		\multirow{2}{*}{\rotatebox{90}{NEW}} &\multicolumn{1}{l}{\stz{{\tt FCRN}/{\tt PESQNet}, EF-intrusive}}   &    {\bf 3.47}  &  \underline{3.87}  & {\bf 0.96}  &    \underline{8.52}  & {\bf 1.95}  &  \underline{3.13}   &  \underline{0.62}  & 7.32 \\ \hhline{~~~~~~~~~~}
		&\multicolumn{1}{l}{\stz{{\tt FCRN}/{\tt PESQNet}, MF-intrusive}}   &    {\bf 3.47}  &  {\bf 3.88}   &    {\bf 0.96}  & {\bf 8.54}  & {\bf 1.95}  &  {\bf 3.18}   &  \underline{0.62}  & {\bf 7.53} \\ \hhline{----------}
	\end{tabular}
	%}
	\label{DNS1_dev}
	\vspace*{-2mm}
\end{table*}
%%%%%%%%%%%%%%%%%%%%%%%%%%%%%55
\vspace*{-1mm}
\section{Experiments and Discussion}
\vspace*{-1mm}
\subsection{Setup, Database, and Metrics}
\vspace*{-1mm}
In this work, signals have a sampling rate of $16\,\text{kHz}$ and we apply a periodic Hann window with frame length of $384$ with a $50\%$ overlap, followed by an FFT with $K=512$. As introduced before, we adopt the {\tt FCRN} proposed in \cite{strake2020fully} as our DNS model. The number of input and output frequency bins in Fig.\,\ref{fig:PESQNet} is set to $K_{\rm in}=260$. The last three frequency bins are redundant for the compatibility with the two maxpooling operations in the employed DNS model from \cite{strake2020fully} and in the proposed {\tt PESQNet} shown in Fig.\,\ref{fig:PESQNet}. For the {\tt PESQNet}, the widths of the employed convolutional kernels are set to $w_i=2^{i-1}$, $i\in\left\{1,2,3,4\right\}$. The number of frames in each feature block shown in Fig.\,\ref{fig:PESQNet} is set to $W=16$.

For the DNS and {\tt PESQNet} pre-training, we employ the same dataset $\mathcal{D}_\text{WSJ0}$ as used in \cite{xu2021deepT}, which is synthesized from the WSJ0 speech corpus \cite{Garofalo2007} clean speech and noise from DEMAND \cite{thiemann2013diverse} and QUT \cite{dean2010qut}. Following \cite{xu2021deepT}, we perform two-stage fine-tuning on the dataset constructed with files randomly chosen from the official Interspeech 2021 DNS Challenge (dubbed DNS3) training material \cite{reddy2021interspeechIN}. This fine-tuning dataset contains $100$ hours of training material $\mathcal{D}^\text{train}_\text{DNS3}$ and $10$ hours of validation material $\mathcal{D}^\text{val}_\text{DNS3}$, employing the same setting used in \cite{xu2021deepT}. The \nth{1}-stage fine-tuning of the DNS and {\tt PESQNet} employs the same loss as used in pre-training, but on $\mathcal{D}^\text{train}_\text{DNS3}$. Afterwards, the DNS and {\tt PESQNet} are fine-tuned jointly on $\mathcal{D}^\text{train}_\text{DNS3}$, utilizing the alternating training protocol shown in Fig.\,\ref{system3}. We use the preliminary synthetic test set from the first Interspeech 2020 DNS Challenge (DNS1) \cite{reddy2020interspeechfinal} (dubbed $\mathcal{D}^\text{dev}_\text{DNS1}$) for development. The final evaluation is reported on the {\it synthetic} test set from the ICASSP 2020 DNS Challenge (DNS2) \cite{reddy2021icassp} (dubbed $\mathcal{D}^\mathrm{test}_\mathrm{DNS2}$).

Following \cite{xu2021deepT}, we employ instrumental metrics such as PESQ \cite{ITUT_pesq_wb_corri}, short-time objective intelligibility (STOI) \cite{taal2010short}, segmental SNR improvement $\Delta\text{SNR}_\text{seg}$ \cite{loizou2013speech}, and speech-to-reverberation modulation energy ratio (SRMR) \cite{falk2010non}. $\Delta\text{SNR}_\text{seg}$ is measured according to \cite{loizou2013speech} to explicitly evaluate the denosing effects on the noisy mixtures without reverberations. SRMR is measured only on the noisy mixtures under reverberated conditions to evaluate the dereverberation effects. Furthermore, we also report the DNSMOS scores \cite{reddy2021dnsmos} on the enhanced speech.
%%%%%%%%%%%%%%%%%%%%%%%%%%%%%%%%%%%%%%%%%%%%%%%%%%%%%%%%%%%%%%%%%%%%%%%%%%
\vspace*{-1mm}
\subsection{Results and Discussion}
\vspace*{-1mm}
In Table\,\ref{DNS1_dev}, we evaluate the performance of the DNS trained with both the EF-intrusive and the MF-intrusive {\tt PESQNet} on the synthetic dataset $\mathcal{D}^\mathrm{dev}_\mathrm{DNS1}$. As baselines, we fine-tune the same pre-trained DNS on $\mathcal{D}^\text{train}_\text{DNS3}$, with either the MSE-based loss \eqref{MT} proposed in \cite{strake2020DNS} (dubbed ``{\tt FCRN} \cite{strake2020DNS}") or with the {\it non-intrusive} {\tt PESQNet} from our previous work \cite{xu2021deepT} (dubbed ``{\tt FCRN}/{\tt PESQNet}, non-intrusive \cite{xu2021deepT}"). Furthermore, we adopt the DNS3 Challenge baseline \cite{braun2020data} as an additional baseline, denoted as ``DNS3 Baseline \cite{braun2020data}". 
%%%%%%%%%%%%%%%%%%%%%%%%%%%%%%%%%%%%%%%%%%%%%%%%%%%%%%%%%%%%%
\begin{table}[t!]
	\centering
	\caption{\textbf{Instrumental quality results} on the \textbf{synthetic test set} $\mathcal{D}^\mathrm{test}_\mathrm{DNS2}$. Best results are in {\bf bold} font, and the second best are \underline{underlined}. Results with marker are depicted in Fig.\,\ref{test_PESQ}.}
	\setlength\tabcolsep{2pt}
	\vspace*{-2mm}
	\begin{tabular}{c c c c c}
		\hline
		& \stz{Method} & PESQ & DNSMOS  &  \stz{STOI}\\ \hline
		\multirow{4}{*}{}& \multicolumn{1}{l}{Noisy} & \stz{2.37} & 3.08 & 0.88\\ \hhline{~~~~~}
		& \multicolumn{1}{l}{\stz{DNS3 Baseline \cite{braun2020data}}} & \stz{3.14} & 3.52 & 0.91\\ \hhline{~~~~~}
		& \multicolumn{1}{l}{\stz{{\tt FCRN} \cite{strake2020DNS}}} & 3.25$^\blacktriangle$ & 3.60 & {\bf 0.93}\\ \hhline{~~~~~}
		& \multicolumn{1}{l}{\stz{{\tt FCRN}/{\tt PESQNet}, non-intrusive \cite{xu2021deepT}}} & \stz{3.34}{\large $^\bullet$} & \underline{3.65} & {\bf 0.93}\\ \hhline{-----}
		\multirow{2}{*}{\rotatebox{90}{NEW}}& \multicolumn{1}{l}{\stz{{\tt FCRN}/{\tt PESQNet}, EF-intrusive}} & \underline{3.36} & \underline{3.65} & {\bf 0.93}\\ \hhline{~~~~~}
		& \multicolumn{1}{l}{\stz{{\tt FCRN}/{\tt PESQNet}, MF-intrusive}} & \stz{\bf 3.37}{\large ${^{\bm \ast}}$} & {\bf 3.67} & {\bf 0.93}\\ \hline
	\end{tabular}
	\label{DNS2_test}
	\vspace*{-3mm}
\end{table}

It can be seen that among all the employed baseline methods, the ``DNS3 Baseline \cite{braun2020data}" performs weakest on PESQ. This could be attributed to the worst noise attenuation reflected by the lowest $\Delta\text{SNR}_\text{seg}$ and the worst dereverberation effects reflected by the lowest SRMR scores. The DNS trained with the {\it non-intrusive} {\tt PESQNet} \cite{xu2021deepT} performs best among all the baselines offering the best or the second-best speech qualities measured by PESQ and DNSMOS under both reverberation conditions. The DNS trained with both versions of the {\it intrusive} {\tt PESQNet} offer comparable quality scores but significantly improves over the DNS3 baseline by about $0.3$ points in terms of PESQ scores under both reverberation conditions. Meanwhile, under the conditions without reverberation, $0.1$ PESQ points improvement is obtained compared to the method trained with MSE-based loss \eqref{MT}, denoted as ``{\tt FCRN} \cite{strake2020DNS}". However, only marginal PESQ improvements ($0.02$ PESQ points) are obtained compared to the method trained with {\it non-intrusive} {\tt PESQNet}. Under the conditions with reverberation, all {\tt PESQNet} methods perform very similar, however, only for SRMR, the MF-intrusive {\tt PESQNet} is clearly the best. Overall, among the DNS trained with the {\it intrusive} {\tt PESQNet}, the MF-intrusive {\tt PESQNet} performs better by offering seven \nth{1}-ranked and one \nth{2}-ranked metrics among all eight employed metrics.
%%%%%%%%%%%%%%%%%%%%
\begin{figure}[t!]
	\centering
	\includegraphics[width=0.47\textwidth]{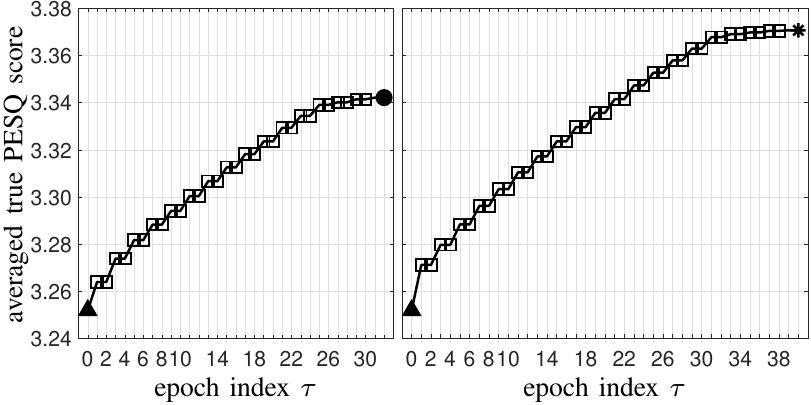}
	\vspace*{-1mm}
	\caption{Averaged true PESQ scores measured on $\mathcal{D}^\text{test}_\text{DNS2}$ {\bf during \mbox{\nth{2}-stage} alternating fine-tuning} with non-intrusive {\tt PESQNet} (left side, \cite{xu2021deepT}) and MF-intrusive {\tt PESQNet} (right side).}
	\label{test_PESQ}
	\vspace*{-2mm}
\end{figure}

In Table\,\ref{DNS2_test}, we measure the instrumental quality on the synthetic test data $\mathcal{D}^\mathrm{test}_\mathrm{DNS2}$. The DNS trained with the MF-intrusive {\tt PESQNet} offers the best performance in all implemented metrics. Compared to the DNS3 baseline and ``{\tt FCRN} \cite{strake2020DNS}", we further increase the PESQ score by $0.23$ and $0.12$ points, respectively. The performance of the DNS trained with both versions of the {\it intrusive} {\tt PESQNet} is very similar. In Fig.\,\ref{test_PESQ}, we plot the averaged true PESQ scores measured on $\mathcal{D}^\mathrm{test}_\mathrm{DNS2}$ during the \nth{2}-stage alternating fine-tuning with {\it non-intrusive} or MF-intrusive {\tt PESQNet}. Compared to the PESQ performance of the DNS before \nth{2}-stage fine-tuning ($3.25$ at $\tau=0$, marker $\blacktriangle$), each epoch of the DNS training mediated by the {\tt PESQNet} (epochs with odd index number) can achieve a better PESQ score until some saturation is reached. The PESQ performance of the DNS trained with the {\it non-intrusive} {\tt PESQNet} converges after epoch $27$ (Fig.\,\ref{test_PESQ}, left side), while the novel MF-intrusive {\tt PESQNet} improves until epoch $36$ (Fig.\,\ref{test_PESQ}, right side). Compared to the use of the {\it non-intrusive} {\tt PESQNet}, there is a slight performance improvement obtained from employing the MF-intrusive {\tt PESQNet} ($0.03$ PESQ points). Accordingly, actually all investigated {\tt PESQNet}s do their job. Note that the {\it non-intrusive} {\tt PESQNet} offers the opportunity to include real training data into the \nth{2}-stage fine-tuning.
%%%%%%%%%%%%%%%%%%%%%%%%%%%%%%
\vspace*{-1mm}
\section{Conclusions}
\vspace*{-1mm}
In this work, we train a deep noise suppression (DNS) model with either a {\it non-intrusive} or an {\it intrusive} {\tt PESQNet}, which is used as a mediator during the DNS training aiming at maximizing the PESQ score of the enhanced speech signal. We investigate two topologies for an {\it intrusive} {\tt PESQNet}, called early-fusion (EF) and middle-fusion (MF) {\tt PESQNet}, respectively. Detailed analyses suggest that the DNS trained with the MF-intrusive {\tt PESQNet} outperforms both the Interspeech 2021 DNS Challenge baseline and the same DNS trained with MSE loss by $0.23$ and $0.12$ PESQ points, respectively. However, compared to the DNS trained with {\it non-intrusive} {\tt PESQNet}, only marginal benefits are obtained, mostly under reverberation. We conclude that it is unnecessary to employ an {\it intrusive} {\tt PESQNet} for DNS training, which opens the possibility to use real training data while achieving comparable performance with employing the still powerful {\it non-intrusive} {\tt PESQNet}.
\clearpage
\bibliographystyle{IEEEbib}
%\ninept
\footnotesize
%\begin{spacing}{0.4}
\bibliography{mainTrans21}
%\end{spacing}
\end{document}